\documentclass{article}
\usepackage{spconf,amsmath,graphicx,amsfonts}
\usepackage{algorithm}
\usepackage{algpseudocode}
\usepackage{underscore}
\usepackage{multirow, cite}
\usepackage{flushend}
\makeatletter

\usepackage{enumitem}
\setlist{nosep, leftmargin=14pt}

\usepackage{mwe} % to get dummy images

\title{DUBLINE: A Deep Unfolding Network for B-line Detection in Lung Ultrasound Images}
\name{\begin{tabular}{c}Tianqi Yang$^{1}$\quad Nantheera Anantrasirichai$^{1}$\quad Oktay~Karakuş$^{2}$\quad Marco Allinovi$^{3}$\\ Hatice Ceylan Koydemir$^{4,5}$\quad Alin~Achim$^{1}$ \end{tabular}\thanks{e-mail: tianqi.yang@bristol.ac.uk}}

\address{$^{1}$ Visual Information Lab, University of Bristol, Bristol, UK\\
        $^{2}$ School of Computer Science and Informatics, Cardiff University, Cardiff, UK\\
        $^{3}$  Nephrology, Dialysis and Transplantation, Careggi University Hospital, Florence, Italy\\
        $^{4}$ Department of Biomedical Engineering, Texas A\&M University, USA\\
        $^{5}$ Center for Remote Health Technologies and Systems, Texas A\&M Engineering Experiment Station, USA}
\begin{document}
%\ninept
\maketitle
\begin{abstract}
In the context of lung ultrasound, the detection of B-lines, which are indicative of interstitial lung disease and pulmonary edema, plays a pivotal role in clinical diagnosis. Current methods still rely on visual inspection by experts. Vision-based automatic B-line detection methods have been developed, but their performance has yet to improve in terms of both accuracy and computational speed. This paper presents a novel approach to posing  B-line detection as an inverse problem via deep unfolding of the Alternating Direction Method of Multipliers (ADMM). It tackles the challenges of data labelling and model training in lung ultrasound image analysis by harnessing the capabilities of deep neural networks and model-based methods. Our objective is to substantially enhance diagnostic accuracy while ensuring efficient real-time capabilities. The results show that the proposed method runs more than 90 times faster than the traditional model-based method and achieves an $F_1$ score that is 10.6\% higher. 
%the proposed method outperforms the traditional model-based method by more than 90 times of efficiency and 10.8\% of $F_1$ score.
\end{abstract}

\section{Introduction}
\label{sec:intro}

Lung ultrasound has emerged as an effective diagnostic tool for various pulmonary conditions. Techniques for analysing lung ultrasound images have been blooming in recent years\cite{yang2022current}. One key finding in lung ultrasound is the presence of B-lines, which appear as laser-like vertical hyperechoic reverberation artefacts originating from the pleural line. B-lines are indicators of the interstitial syndrome and are useful for the diagnosis of conditions like pneumonia and pulmonary edema, and have also been shown to correlate with the volume of extravascular lung water \cite{allinovi2017lung}. Thus, accurate detection of B-lines allows non-invasive evaluation of pulmonary edema and other conditions involving interstitial syndromes. 

\begin{figure}[t!]
\centering
\includegraphics[width=1\linewidth]{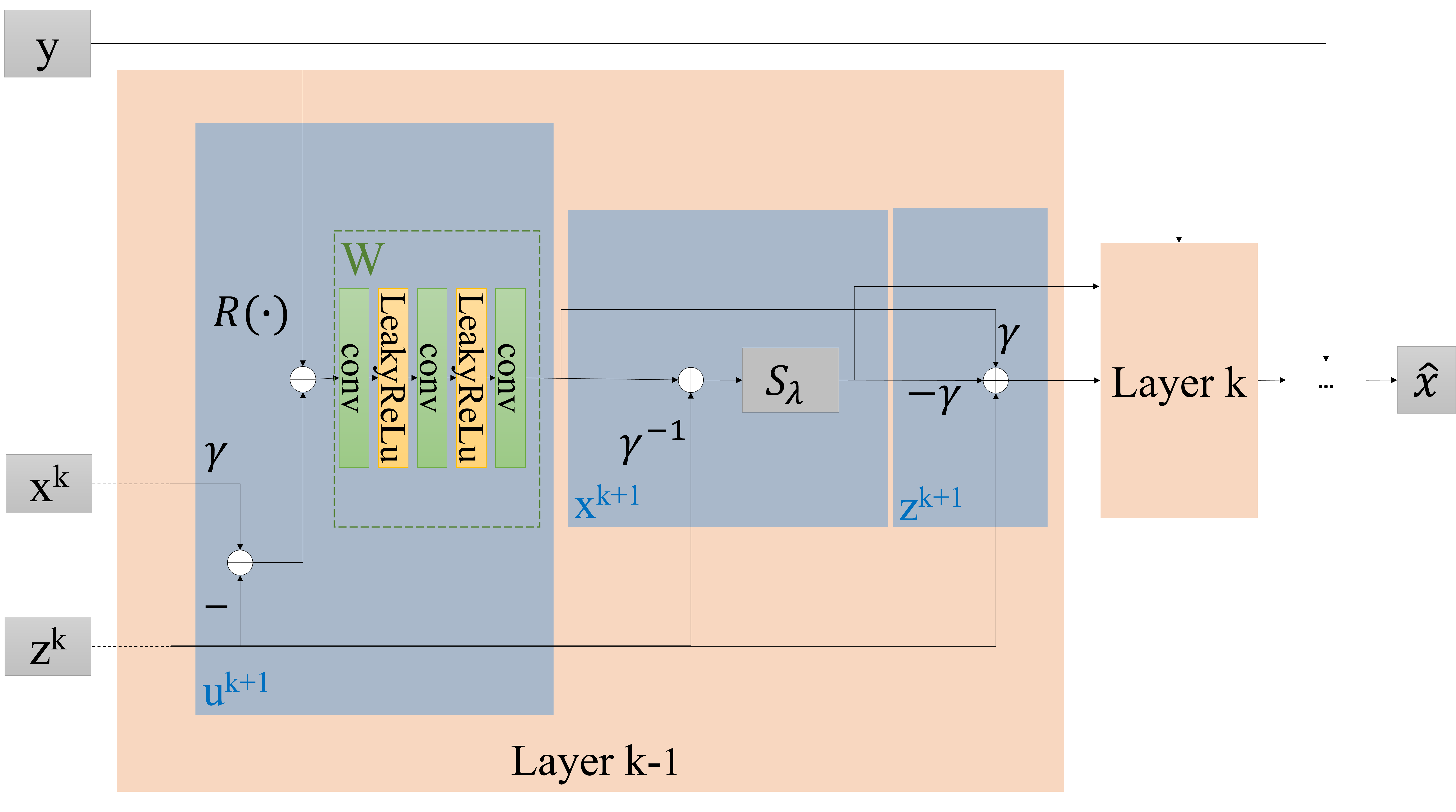}
\caption{Diagrammatic representation of DUBLINE. Trainable parameters are coloured in green. }
\label{fig:ADMM-net}
\vspace{-5mm}
\end{figure}

In previous efforts to automatically detect B-lines, various techniques were employed. Hand-crafted image processing techniques, like polar reformatting and thresholding, were used by Brattain et al.  \cite{brattain2013automated}. Anantrasirichai et. al. first posed line detection as an inverse problem \cite{anantrasirichai2016automatic,anantrasirichai2017line}, where a B-mode LUS image is converted to a representation of radius and orientation in the Radon domain. The inverse problem is then solved using the alternating direction method of multipliers (ADMM) \cite{boyd2011distributed}. The method was further developed for evaluating COVID-19 patients by Karakus et. al. \cite{karakucs2020detection}, whereby improved performance was achieved by regularising the solution using the Cauchy proximal splitting (CPS) algorithm \cite{karakucs2020convergence}.

With the advent of deep learning, convolutional neural networks (CNNs) have been applied for robust B-line detection. Van Sloun and Demi \cite{van2019localizing} applied a CNN with gradient-based class activation mapping (Grad-CAM) \cite{selvaraju2017grad} for B-line detection in single LUS frames. Alternatively, Baloescu et. al. \cite{baloescu2020automated} used a relatively shallow custom-made model architecture with 3-D filters (3-D CsNet) for B-line assessment in a supervised manner. To extract better features across the video, Kerdegari et. al. \cite{kerdegari2021b} combined the long-short-term memory (LSTM) network and spatiotemporal attention mechanism to achieve B-line localisation. Our own work in \cite{yang2023semi} employed contrastive learning to investigate the representation of B-lines in an unsupervised manner. This was followed by fine-tuning on a limited labelled dataset, resulting in a significant reduction in the need for manual annotations.

In this work, we propose to apply the deep unfolding technique to B-line detection in lung ultrasound. This aims to achieve a totally unsupervised real-time B-line detection. The design of the DUBLINE network was inspired by \cite{anantrasirichai2016automatic}, and the architecture is shown in Fig.\ref{fig:ADMM-net}, where a CNN is used to replace the complex Radon transform computations in the variable update steps. We compare the performance with the method in \cite{anantrasirichai2017line} following the same detection process, demonstrating superiority in terms of precision, recall, accuracy, $F_1$ score, and time efficiency.

\section{Methodology}
\label{sec:method}

\subsection{Line artefact model}
\label{sec:inverse}
The line artefact in the noisy ultrasound images can be modelled in terms of its inverse Radon transform as 
\begin{equation}\label{model}
y=R^{-1}x+n,
\end{equation}
where $y$ is the observed ultrasound image, $x$ is the line represented by a distance $r$ from the centre of $y$ and a orientation $\omega$ from the horizontal axis of the image. $R$ and $R^{-1}$ represent the Radon transform and its inverse, respectively whilst $n$ refers to the additive noise. In a general formulation without the noise, the Radon transform is described as in Eq.\ref{radon}, where \(\delta(\bullet)\) is the delta function. 
\begin{equation}\label{radon}
x=\int_{\mathbb{R}^{2}}y(i,j)\delta (r-icos\omega -jsin\omega )didj.
\end{equation}
Using $l_1$ regularisation, the estimation of lines can be found by solving the following optimisation problem with $\alpha$ being the regularisation constant: 
\begin{equation}\label{opt}
\hat{x}=\arg \min_{x} \left \{ \frac{1}{2}\left \| y-R^{-1}x \right \|_{2}^{2} +\alpha \left \| x \right \|_{1}\right \}.
\end{equation}
where discrete operators $R$ and $R^{-1}$. % are used as in \cite{kelley1993fast}. 

\subsection{Optimisation problem}
\label{sec:inverse}
The problem in Eq.\ref{opt} can be solved by employing ADMM \cite{boyd2011distributed}, which is a variant of the augmented Lagrangian scheme that uses partial updates for the dual variables, so that Eq.\ref{opt} is equivalent to
\begin{equation}\label{ADMM}
\begin{aligned}
\mathrm{minimize}&\quad \frac{1}{2}\left \| y-R^{-1}u \right \|_{2}^{2} +\alpha \left \| x \right \|_{1},\\
\mathrm{subject\; to}&\quad u-x=0.
\end{aligned}
\end{equation}
% \begin{equation}\label{ADMM}
% \begin{aligned}
% \mathrm{minimize}&\quad  f(u) + g(x),\\
% \mathrm{subject\; to}&\quad u-x=0.
% \end{aligned}
% \end{equation}
% where
% \begin{subequations}
% \begin{align}
% f(u)&=\frac{1}{2}\left \|  y-R^{-1}u\right \|_{2}^{2}, \label{fu}\\
% g(x)&=\alpha \left \| x \right \|_{1}. \label{gx}
% \end{align}
% \end{subequations}
Then the augmented Lagrangian for Eq.\ref{ADMM} is 
\begin{equation}\label{Lagrangian}
\begin{aligned}
\mathcal{L}_{\gamma}=\frac{1}{2}\left \|  y-R^{-1}u\right \|_{2}^{2}+\alpha \left \| x \right \|_{1}+z^{T}(u-x)+\frac{\gamma }{2}\left \|  u-x\right \|_{2}^{2},
\end{aligned}
\end{equation}
where $z$ is the Lagrange multiplier, and $z^{T}$ indicates the transpose of $z$. For penalty parameter $\gamma>0$, the optimisation problem in Eq.\ref{ADMM} can be solved by the iterative scheme of ADMM as stated in Eq.\ref{unfoldu} to \ref{unfoldz}.

In traditional model-based methods \cite{anantrasirichai2016automatic,anantrasirichai2017line,karakucs2020detection}, optimisation can be slow for real-time applications. Deep unfolding, introduced by Gregor and LeCun \cite{gregor2010learning}, combines unsupervised learning and rapid convergence by training a feedforward neural network to approximate iterative algorithms like ADMM.\\
\textbf{Neural network formulation:} The three-step iteration in Eq.\ref{unfoldu} to \ref{unfoldz} can be formulated as a neural network. The corresponding unfolded network is depicted in Fig.\ref{fig:ADMM-net}. The number of layers corresponds to the iteration counter in the traditional ADMM algorithm.

For updating $u$, the problem is solved as
\begin{equation}\label{unfoldu}
\begin{aligned}
u^{k+1}&=\arg \min_{u}\frac{1}{2}\left \|  y-R^{-1}u\right \|_{2}^{2}+(z^{k})^{T}(u-x^{k})\\& \quad+\frac{\gamma }{2}\left \|  u-x^{k}\right \|_{2}^{2},\\&=((R^{-1})^{T}R^{-1}+\gamma I)^{-1}((R^{-1})^{T}y+\gamma x^{k}-z^{k}),\\&=W_{\theta }(y, x^{k}, z^{k}).
\end{aligned}
\end{equation}
where $k$ is an internal iteration counter, $I$ denotes the identity matrix, $W_{\theta}$ denotes a CNN with 3 convolutional layers and LeakyReLu being the activation function. $\theta$ represents its parameters. In this paper, we use \textit{radon\_transformation} functions$\footnote{https://github.com/drgHannah/Radon-Transformation/tree/main}$ to construct $R$ and $R^{-1}$. $(R^{-1})^{T}$ serves as the forward Radon projection.%\cite{gengsheng2010medical}.

For updating $x$, we have
\begin{equation}\label{unfoldx}
\begin{aligned}
x^{k+1}&=\arg \min_{x}\frac{\alpha}{\gamma }\left \| x \right \|_1+\frac{1}{2}\left \| u^{k+1}-x+\frac{z^{k}}{\gamma } \right \|_{2}^{2},\\&=\mathit{S}_{\lambda }(u^{k+1}+\frac{z^{k}}{\gamma }).
\end{aligned}
\end{equation}
where $\mathit{S}_{\lambda }(\bullet)$ is a soft thresholding described as
\begin{equation}\label{thresholding}
\begin{aligned}
\mathit{S}_{\lambda }(a)=sign(a)\mathrm{max}(\left | a \right |-\lambda ,0),
\end{aligned}
\end{equation}
In this work, we take the maximum absolute row sum as the threshold. Lastly, each iteration for updating $z$ follows%Eq.\ref{z}.\\
\begin{equation}\label{unfoldz}
z^{k+1}= z^{k}+\gamma(u^{k+1}-x^{k+1}).
\end{equation}
\textbf{Loss function:}  
% As we specifically interested the line structures in the lung ultrasound images, we implemented the radon transform within the angle $\Theta_{v}\in\left \{ \pm15^{\circ} \right \}$ ($0^{\circ}$ starts from the $x$ axis) for vertical lines and $\Theta_{h}\in\left \{90^{\circ} \pm5^{\circ} \right \}$ for horizontal lines. 
We train the network to maximise the structural similarity index measure (SSIM) %\cite{wang2004image} 
between the reconstructed line structures and the input image, and therefore the loss function is
\begin{equation}\label{loss}
\begin{aligned}
\mathit{L}= (1-SSIM(R^{-1}\hat{x},y)).
\end{aligned}
\end{equation}
%where the $SSIM$ is defined as
%\begin{equation}\label{ssim}
%\begin{aligned}
%SSIM(a,b)=\frac{(2\mu_{a}\mu_{a}+c_{1})(2\sigma _{ab}+c_{2})}{(\mu _{a}^{2}+\mu _{b}^{2}+c_{1})(\sigma_{a}^{2}+\sigma_{b}^{2}+c_{2})}.
%\end{aligned}
%\end{equation}
%The $\mu$ is the pixel sample mean of the image, $\sigma^{2}$ is the variance of the image and $\sigma _{ab}$ is the covariance of the images. $c_{1}$ and $c_{2}$ are two variables to stabilize the division with a weak denominator.

Considering the nature of our problem, instead of other practical loss functions like MSE (regression), or cross-entropy (classification), we choose to promote SSIM thanks to its (i) artefact reduction capabilities, and (ii) robustness to common image degradations such as noise, or blurring. Ultrasound imagery and B-line detection are both suitable for the utilisation of SSIM due to aforementioned reasons and we leave the exploration of other loss functions for future work.

\subsection{B-line detection}
\label{sec:detection}
The B-line detection procedure in this paper aligns with \cite{anantrasirichai2017line}, which detects the lines in the Radon domain using the local maxima technique. To reduce the influence of skin and muscle, the top part of the image is dimmed. %\cite{lowe2004distinctive}. 
The procedure starts with detecting the pleural line in the restored $\hat{x}$ within the angle $\Theta_{h}\in \left [ 90^{\circ} \pm5^{\circ}\right ]$ ($0^{\circ}$ starts from the $x$ axis). Then, the A-lines - physiological horizontal lines below the pleural line - are detected by using the same range of angles. The B-lines are detected at $\Theta_{v}\in\left [ \pm15^{\circ}\right ]$. Any vertical artefacts over-passed by an A-line are removed and all detected vertical lines that originate from the same pleural line point are counted as a single B-line.
% The complete training procedure is depicted in Algorithm \ref{Algorithm}.
% \begin{algorithm}
% \caption{Radon-ADMM Training Procedure}\label{Algorithm}
% \begin{algorithmic}[1]
% \State \textbf{Input:} $y$, num_iter, epoch
% \State \textbf{Initialization:} $u^{0}$, $x^{0}$, $z^{0}$, $k=0$\\
% $x_{v}=\mathit{R}_{\Theta_{v}}(y)$, $x_{h}=\mathit{R}_{\Theta_{h}}(y)$

% \While{\textit{$k<num\_iter$}}
%     \State \textit{$u_{v}^{k+1}=W_{\theta }(y, x^{k}, z^{k})$}
%     \State \textit{$x^{k+1}=\mathit{S}_{\lambda }(u^{k+1}+\frac{z^{k}}{\gamma })$}
%     \State \textit{$z^{k+1}=z^{k}+\gamma(u^{k+1}-x^{k+1})$}
% \EndWhile

% \State \textbf{Output:} $y$
% \end{algorithmic}
% \end{algorithm}
\section{Experimental Results}
\label{sec:setup}
\begin{figure}[t!]
\centering
\includegraphics[width=1\linewidth]{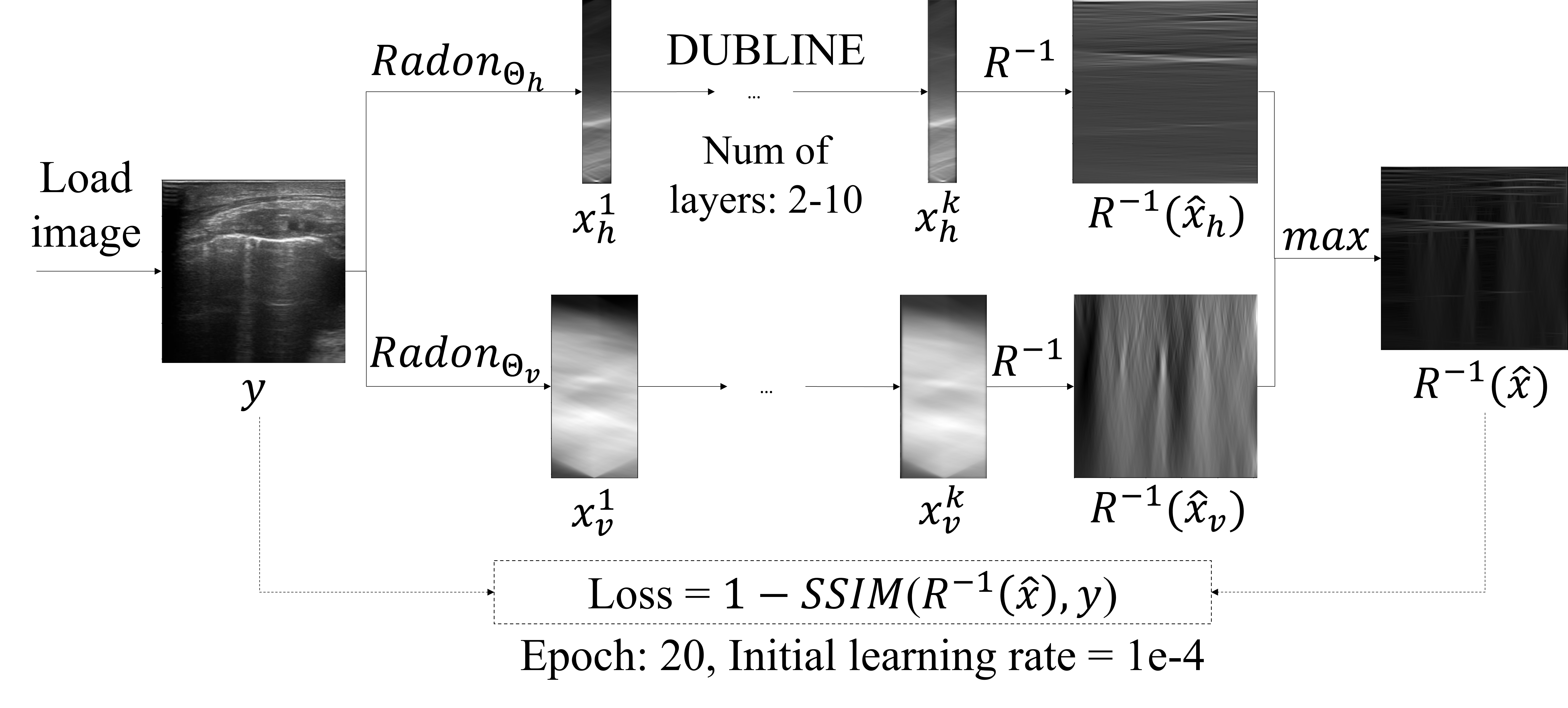}
\caption{Training procedure of DUBLINE.}
\label{fig:train}
\vspace{-5mm}
\end{figure}

\textbf{Dataset and pre-processing:} The data was obtained from the Nephrology, Dialysis and Transplantation, Careggi University Hospital, Florence, Italy. We trained the network on unlabeled lung ultrasound
images. We used 1000 images as our training set and 122 images as the test set, from 34 patients. LUS evaluations were performed whilst patients attended for regular haemodialysis, using an ultrasound machine (MyLab Class C-Esaote$^{\circledR}$, Genoa, Italy) with a 6–18 MHz linear probe. There are in total 161 B-lines included in the test set and the ground truth was provided by a physician with long-term expertise in LUS. All the images were resized to 256 $\times$ 256 pixels.\\
% We trained our network on 80 unlabelled lung ultrasound images and tested the model on 122 lung ultrasound images acquired from from Paediatric Nephrology Unit, Meyer Childrens Hospital, Florence, Italy. 
\textbf{Experimental setup:} As we are specifically interested in the line structures in the lung ultrasound images, in DUBLINE, we implemented the Radon transform with the angle $\Theta_{v}$ and $\Theta_{h}$ as stated in the previous section. We set $\alpha$ and $\gamma$ equal to 1 and the initial learning rate to $10^{-4}$ after a trial-error step. The learning rate decays every 10 epochs. The output of the network is acquired by comparing the reconstructed images and taking the maximum value of each pixel. We also evaluated DUBLINE for the number of layers of the unfolded network ranging from 2 to 10. The networks are all trained for 20 epochs. The training procedure is depicted in Fig.\ref{fig:train}. The network is implemented in PyTorch and trained on Nvidia Geforce RTX 3090. The method in \cite{anantrasirichai2017line} is reproduced using 12th Gen Intel$^{\circledR}$ Core\textsuperscript{TM}  i7-12700 2.10 GHz.\\
\textbf{Results Analysis:} We compared the proposed method with the method in \cite{anantrasirichai2017line}, and adjusted the parameters accordingly to improve the detection performance. Our assessment focused on various performance metrics, including true positives (TP), false positives (FP), false negatives (FN), precision, recall, and the $F_{1}$ score. The results of this comparison are summarized in Table \ref{compare}, where ``DU" denotes our proposed DUBLINE method. When assessing the model's performance across different numbers of iterations, it becomes evident that training DUBLINE with more than 2 iterations (equivalent to employing more than 2 layers of DUBLINE) consistently outperforms the baseline method \cite{anantrasirichai2017line} across all the metrics. This can be attributed to the traditional method typically requiring more than 2 iterations in the ADMM algorithm to meet the stopping criteria. As a result, the deep unfolded network necessitates at least 3 layers to achieve equal or superior performance to the method outlined in \cite{anantrasirichai2017line}. Further examination of iterations 3 through 10 reveals that DUBLINE exhibits higher TP rates and lower FN rates, leading to a significant improvement of at least 10.5\% in recall for the detection task. The best-performing iteration is observed at iter8, which yields the highest TP count (100) and the lowest FN count (61), resulting in a notable 10.6\% enhancement in the $F_{1}$ score. Across the board, the number of FPs is reduced, with the most substantial reduction being 13 when compared to the baseline. While there are exceptions at iter3 and iter9, the precision scores are still improved by 4.9\% and 6.4\%, respectively. 

\begin{figure}[t!]
\centering
\includegraphics[width=1\linewidth]{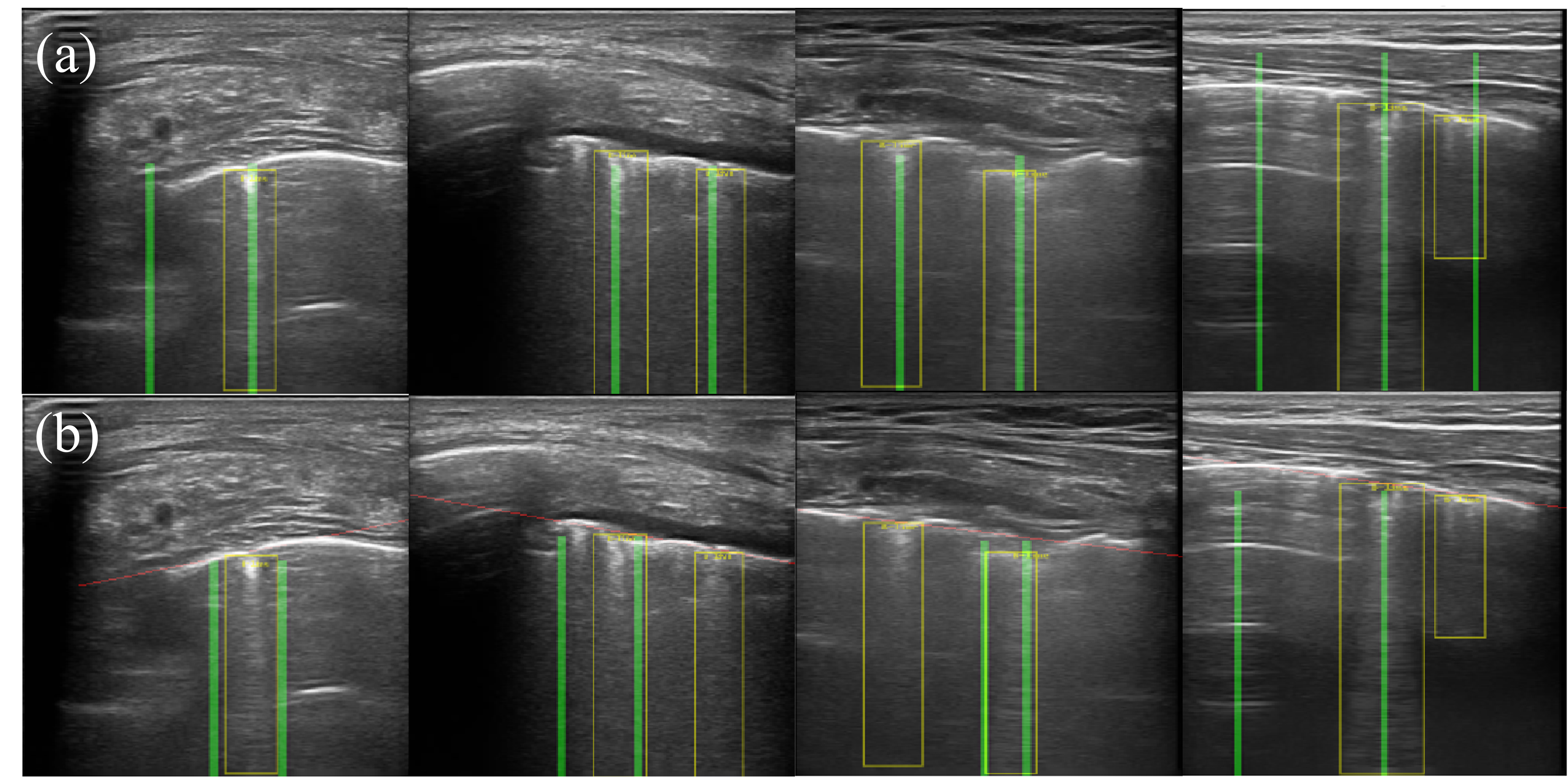}
\caption{B-line detection results. (a) is the results of the DUBLINE and (b) is the results of \cite{anantrasirichai2017line}. The yellow boxes are the bounding boxes of the ground truth. The green lines are the detected B-lines.}
\label{fig:detection}
\vspace{-5mm}
\end{figure}

Fig.\ref{fig:detection} shows some examples of B-line detection results. When the green lines fall within half of the ground truth box width centring at the centre line of the ground truth box, they are counted as TP. If the green lines fall outside of this range, they are counted as FP. When there is no green line within the yellow box, it is counted as an FN. As shown in the examples, compared to the method described in \cite{anantrasirichai2017line}, the proposed method detects more correct B-lines and significantly improves the accuracy of their detected positions. However, the proposed method exhibits relatively poor differentiation between the true B-lines and the bright vertical lines that only look similar to B-lines, causing FPs to often occur at the positions of bright vertical artefacts. In cases when the pleural line is darker than the horizontal artefacts above it, the detected position of the pleural line is likely to be biased, resulting in the starting point of the B-lines being higher than the actual position.

We highlight the time efficiency of the DUBLINE by comparing it with that of the method in \cite{anantrasirichai2017line}. The execution time per image of the DUBLINE is only 0.0186 seconds, whereas the average speed of the traditional ADMM is 1.6803 seconds per image. The unfolded algorithm significantly improves the speed by more than 90 times with less than 0.5 seconds to process 24 frames in a real-time video. 

\begin{table}[t!]
\caption{Detection results from different settings}
\small
\begin{tabular}{|cc|c|c|c|c|c|c|}
\hline
\multicolumn{2}{|c|}{}                                                                                                           & TP  & FP           & FN & Precision      & Recall         & F1             \\ \hline
\multicolumn{2}{|c|}{\cite{anantrasirichai2017line}}                                                                                                      & 74  & \textbf{166} & 87 & 0.308          & 0.46           & 0.369          \\ \hline
\multicolumn{1}{|c|}{\multirow{9}{*}{\begin{tabular}[c]{@{}c@{}}DU\end{tabular}}} & iter2  & 61  & 154          & 100 & 0.284          & 0.379          & 0.324          \\ \cline{2-8} 
\multicolumn{1}{|c|}{}                                                                                                  & iter3  & 94 & 169          & 67 & 0.357          & 0.584          & 0.443          \\ \cline{2-8} 
\multicolumn{1}{|c|}{}                                                                                                  & iter4  & 93 & 158          & 68 & 0.371          & 0.578          & 0.451          \\ \cline{2-8} 
\multicolumn{1}{|c|}{}                                                                                                  & iter5  & 91 & 157          & 70 & 0.367          & 0.565          & 0.445          \\ \cline{2-8} 
\multicolumn{1}{|c|}{}                                                                                                  & iter6  & 97 & 158          & 64 & 0.38          & 0.602          & 0.466          \\ \cline{2-8} 
\multicolumn{1}{|c|}{}                                                                                                  & iter7  & 93 & \textbf{153}          & 68 & 0.378          & 0.578          & 0.457          \\ \cline{2-8} 
\multicolumn{1}{|c|}{}                                                                                                  & iter8  & \textbf{100} & 160          & \textbf{61} & \textbf{0.385}          & \textbf{0.621}          & \textbf{0.475}          \\ \cline{2-8} 
\multicolumn{1}{|c|}{}                                                                                                  & iter9  & 99 & 167          & 62 & 0.372           & 0.615          & 0.464          \\ \cline{2-8} 
\multicolumn{1}{|c|}{}                                                                                                  & iter10 & 98 & 162          & 63 & 0.377 & 0.609          & 0.466 \\ \hline
\end{tabular}
\label{compare} 
\end{table}

\section{Conclusion}
\label{sec:conclusion}
In this paper, we propose a fast and lightweight method based on a deep unfolded ADMM algorithm to tackle B-line detection in lung ultrasound as an inverse problem, with the ultimate goal of achieving completely unsupervised real-time B-line detection. Compared to the traditional model-based method, the proposed approach shows an improvement in precision, recall, accuracy, $F_{1}$ score, and time efficiency. However, the limitation still exists in the detection procedure. Future works will explore possible unsupervised approaches that can further enhance the accuracy and robustness of B-line detection. 
\section{Ethical Statement}
\label{sec:ethic}
The study protocol conformed to the Declaration of Helsinki and was approved by a local research ethics committee (study approval number 18217/OSS). Informed consent was obtained from all subjects involved in the study.
\small
\bibliographystyle{IEEEtran}
\bibliography{ref}

\end{document}